\documentclass[fleqn]{olplainarticle}
\usepackage[utf8]{inputenc}
\usepackage[
    backend=biber,
    style=authoryear-icomp,
    sortlocale=UTF-8,
    natbib=true,
    url=true,
    doi=true,
    eprint=false
]{biblatex}
\PassOptionsToPackage{hyphens}{url}
\RequirePackage{hyperref}
\hypersetup{colorlinks,allcolors=blue}
\usepackage{csquotes}
\usepackage{etoolbox,xpatch}
\usepackage{graphicx}

\newtheorem{question}{Question}

\title{Standardizing Representation for Equality with a Population Seat Index}

\author[1]{Liang Zhao}
\author[1]{Akiko Tanimoto}
\author[1]{Wenruo Lyu}

\affil[1]{Graduate School of Advanced Integrated Studies in Human Survivability (Shishu-Kan),
Kyoto University, Kyoto, Japan. Email: L. Zhao \url{liangzhao@acm.org}}


\begin{abstract}
	{\em Proportional representation} (PR) has long been believed the ideal system
	for {\em the equality of individuals} in apportioning the seats of a legislature body
	to subgroups.
	We observe that 
	PR implicitly assumes the (standard) number of representatives
	is proportional to the population, {\em a situation no longer observed since 1820s}.
	To address this issue, we suggest to formulate the apportionment problem in a
	broader context by {\em explicitly} specifying a standard function $f$
	such that $f(p)$ is the standard, possibly fractional number of representatives for population $p$,
	where PR assumes $f(p)\propto p$.
	For this {\em generalized apportionment problem},
	we give a {\em population seat index} (PSI) $\frac{f^{-1}(s)}{p}$ for quantifying
	the contribution of an individual in assigning $s$ seats to
	a population $p$, where $f^{-1}$ is the inverse of $f$.
	With the PSI, we derive apportioning schemes with absolute and relative individual equality.
	Particularly, for $s$ seats, populations $p_1, \ldots, p_k$, and a standard function
	$f(p) = a + b p^\gamma$ with constants $a, b, \gamma \ge 0$,
	the ideal, possibly fractional number of seats for subgroup $i$ is
	$a + \frac{(S-ka)p_i^{\gamma}}{\sum p_j^{\gamma}}$,
	not $\frac{Sp_i}{\sum p_j}$ calculated by PR which works only for $a=0$, $\gamma=1$.
	Finally, since real-world observations indicate a standard function $f \propto p^\gamma$ with $\gamma < 1$,
	we conclude that PR represents individuals in less populous subgroups {\em less}
	than individuals in more populous subgroups.
\end{abstract}

\begin{document}

\maketitle

\section{Introduction}
\label{sec:introduction}

	%
	%
	{\em Proportional representation} (PR) is a core principle in law and political studies.
	As expressed by the slogan ``one person, one vote'',
	PR asks to reflect subgroups of a
	population (or an electorate or votes; in the following work, we use ``population'' for
	ease of writing.) {\em proportionally} in a legislature or an elected body.
	This proportionality is believed the ideal system for ensuring the equality of individuals
	(see, e.g., \cite{lijphart_1998}).
	PR has been adopted worldwide in modern politics, apportionment, and elections
	(see, e.g., \cite{allen_2017,benoit_2000,lijphart_1998,lijphart_2012,pukelsheim_2017,puyenbroeck_2008,samuels_2001,taagepera_2003,pr}).

	The core of PR is the {\em proportion of seats to population} (PSP),
	used to calculate the contribution (or weight) of an individual in distributing seats.
	For example, in apportioning the seats of a legislative to subgroups,
	the PSP of a subgroup $i$ is defined as $\frac{s_i}{p_i}$, where $s_i$ and $p_i$ denote
	the number of distributed seats and the population of subgroup $i$ respectively.
	PR requires $\frac{s_i}{p_i} = \frac{s_j}{p_j}$ for any $i$ and $j$.
	An apportionment not satisfying this requirement is considered to violate
	the representation equality of individuals
	(\cite{auerbach_1964,eckman_2021,frederick_2008,huntington_1942,samuels_2001}).

	Unfortunately, we observe that the PSP has a bias in calculating the contribution of an individual.
	Speaking precisely, the PSP implicitly assumes the (standard) number of representatives
	is proportional to the population, {\em a situation no longer observed since 1820s}.
	This observation implies that PR is {\em insufficient} for ensuring individual representation equality.
	%
	%
	%
	%
	Let us explain the observation with the following problem.

	\begin{question}[Equal contribution problem]\label{que:contribution}
		Suppose that a group of $p>0$ people contributed equally to an outcome $s>0$.
		What is the contribution of an individual within that group?
	\end{question}
	
	A simple answer is the proportion $\frac{s}{p}$, i.e., the PSP.
	This {\em per capita} measure, however, is known to be limited.
	For example, on the well-known {\em per capita gross domestic product} (per capita GDP),
	West pointed out that ``simple linear proportionality, implicit in using per capita measures,
	is almost {\em never} valid'' (\cite{west_2017}, p18).
	We illustrate the limitation of the per capita measure $\frac{s}{p}$ with some examples.
	First, assume that $s$ is independent of $p$, the model of the US Senate.
	In this case, the value of the contribution of an individual should be {\em zero},
	since an individual contributes nothing to the outcome $s$ by the assumption
	that $s$ is independent of $p$.
	However, the per capita measure $\frac{s}{p} \neq 0$.

\begin{figure}[htbp]
		\centering
		\includegraphics[width=0.5\textwidth]{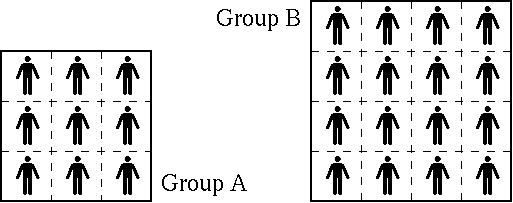}
		\caption{An illustration on the bias of the per capita measure $\frac{s}{p}$.
		Let the outcome $s=\sqrt{p}$ be the $\frac{1}{4}$ of the perimeter of the square by $p$ people.
		Then $\frac{s}{p}$ estimates the contribution of an individual of Group A as
		$\frac{3}{9}=\frac{1}{3}$, larger than that of Group B which is $\frac{4}{16}=\frac{1}{4}$,
		although all people are completely identical!}
		\label{fig:illustration}
	\end{figure}

	Next, assume that the outcome $s$ is proportional to the $\gamma^{\rm th}$-power of the number of people,
	i.e., $s \propto p^\gamma$, with a constant $\gamma < 1$.
	Then, $\frac{s}{p} \propto p^{\gamma-1}$, estimating the contribution
	of an individual in a larger group {\em less than}
	a smaller group (see an illustration in Figure~\ref{fig:illustration}).
	On the other hand, if $\gamma > 1$, then $\frac{s}{p} \propto p^{\gamma-1}$, estimating
	the contribution of an individual in a larger group {\em more than} a smaller group.
	This population-dependent bias does not exist if and only if $s \propto p$.
	%
%
	
	The above analysis implies that {\em the per capita measure $\frac{s}{p}$ such as the PSP
	can correctly estimate
	the contribution of an individual only for such a special situation that the outcome $s$ is proportional
	to the population $p$} (\cite{west_2017}).
	Therefore, PR is limited, as it relies on the population-dependent index PSP.
	To calculate individual contribution (for ensuring equality),
	an apportioning scheme depends
	on the assumption on the standard number of representatives (i.e., the outcome) for a population.
	The PSP, hence the PR, lacks formal correctness as it implicitly assumes the proportionality
	between the number of representatives and the population which, as discussed later,
	is no longer observed since 1820s.

	Therefore, to apportion seats to subgroups equally, we first need to know an ideal or standard
	number of representatives for a subgroup (to define the equality).
	If there is no evidence showing the existence of such a number,
	then PR seems the only choice---not because it is correct but because there is no better option.
	However, today we {\em have} quite a large literature on an ideal number of representatives
	for a population.
	It makes possible to standardize the representation {\em now}.
	
	In this paper, we propose to formulate the apportionment problem in a
	broader context by {\em explicitly} specifying a standard function $f$
	such that $f(p)$ is the standard, possibly fractional number of representatives for population $p$,
	where PR assumes $f(p)\propto p$.
	For this {\em generalized apportionment problem}, we provide a {\em population seat index} (PSI)
	to calculate the unbiased contribution (weight) of an individual and uses it to obtain
	apportionment with absolute and relative representation equality.
	It is shown that PR {\em overestimates} (respectively, {\em underestimates}) the contribution
	of individuals in a less (more) populous groups, thus resulting in {\em underrepresentation}
	({\em overrepresentation}) for the people within those groups.
	In contrast, the proposed scheme guarantees the representation equality of individuals.

\section{Literature review}\label{sec:review}

	Determining an ideal number of representatives (seats) for a population is
	the most ``susceptible'' political problem (\cite{madison_1788}).
	This problem was divided into two subproblems in the 1920s {\em for ease of operation}
	(\cite{chafee_1929}).
	One subproblem is determining an optimal size of a legislative body,
	i.e., the {\em total} number of seats.
	Another subproblem is optimally apportioning a {\em fixed} number of seats to subgroups.
	These two subproblems were originally linked by the Framers of the U.S. Constitution
	(\cite{madison_1789}), but unlinked by Title 2 of the U.S. Code (\cite{kromkowski_1991}).
	However, as discussed in Section~\ref{sec:introduction},
	they are indeed {\em dependent} and cannot be considered separately:
	apportioning seats requires to assume the standard number of representatives for a population.
	In this section, we review the literature on both subproblems and relink them in the next section.

\subsection{Standard number of representatives for a population}\label{subsec:number}

	There are many studies on an optimal number $s^*=f(p)$ of representatives
	(seats) for a population $p$.
	Notice that $s^*=f(p)$ assumes the equality of individuals, since it depends
	only on the population, not on any individual property.
	Table~\ref{tab:number} summarizes the key results, as far as the authors know.
	All of these results can be formulated or approximated
	by Formula~\ref{eqn:general} with different values of $\gamma$.
	%
	Notice that these studies cover both national and subnational legislatures with
	empirical and theoretical analysis, suggesting the existence of an ideal size of
	legislatures for a population.
	\begin{equation}\label{eqn:general}
		s^* \ = \ f(p) \ \propto \ {p}^{\gamma} \ \mbox{ for some constant $\gamma$ with $0 \le \gamma \le 1$}.
	\end{equation}

	\begin{table}[htb]
		\caption{Key studies on a standard number of representatives (see
		Formula~\ref{eqn:general} for the meaning of $\gamma$).}\label{tab:number}
	\centering
	\begin{tabular}{c c c l}
		{Scheme} & {$\gamma$} & {Reference} & {Data/Model/Remark} \\\hline\hline
		Empirical &   $\approx 0.37$    & \cite{stigler_1976}  & parliaments of 37 democratic countries\\
		(Regression) &   $\approx 0.41$    & \cite{auriol_2012}  & parliaments of 111 countries\\
		&   $\approx 0.39$    & \cite{zhao_2020}  & parliaments of 192 countries\\
		&   $\approx 0.23$ & \cite{stigler_1976} & 49 lower houses of states in the US\\
		&   0.24--0.55 & \cite{tanimoto_2022} & subnational legislatures of 9 countries \\
		&   $\approx 0.37$    & Figure~\ref{fig:us}, this article & U.S. Congress between 1790 and 1920 \\ 
		&      $\approx 1$    & Figure~\ref{fig:us_house}, this article  & U.S. House before 1830 \\ \hline \hline
		Theoretical &   &  & \\\hline
		Fixed-size  &      $0$    & U.S. Senate  &  100 (two seats per state)\\
		&      & U.S. House after the 1920s  & 435 \\\hline
		Cubic root    &      $1/3$    & \cite{taagepera_1972}  & Social mobilization model\\\hline
		Sublinear    &      $\frac{1}{3} \le \gamma \le \frac{5}{9}$    & \cite{zhao_2020}  & Social network model\\
		&      (e.g.) $0.4$    & (Same as above)  & First model matching real-world data  \\\hline
		Square root   &    $1/2$      & \cite{penrose_1946}  & Voting power model\\
		&      & \cite{auriol_2012}  & Mechanism design\\
		& & \cite{godefroy_2018}  & Mechanism design\\
		&      & \cite{gamberi_2021}  &  Complex network model\\
		&      & \cite{margaritondo_2021}  &  Revision of \cite{taagepera_1972}\\
		&      & \cite{blonder_2021}& Derived from \cite{madison_1789}\\\hline
		Proportional     &      $1$ &  \cite{ismail_2018}  & Voting model (fixed voting cost)\\
		 &       & \cite{revel_2022}  & Voting model \\\hline
	\end{tabular}
	\end{table}

\begin{figure}[htbp]
		\centering
		 \includegraphics[width=\textwidth]{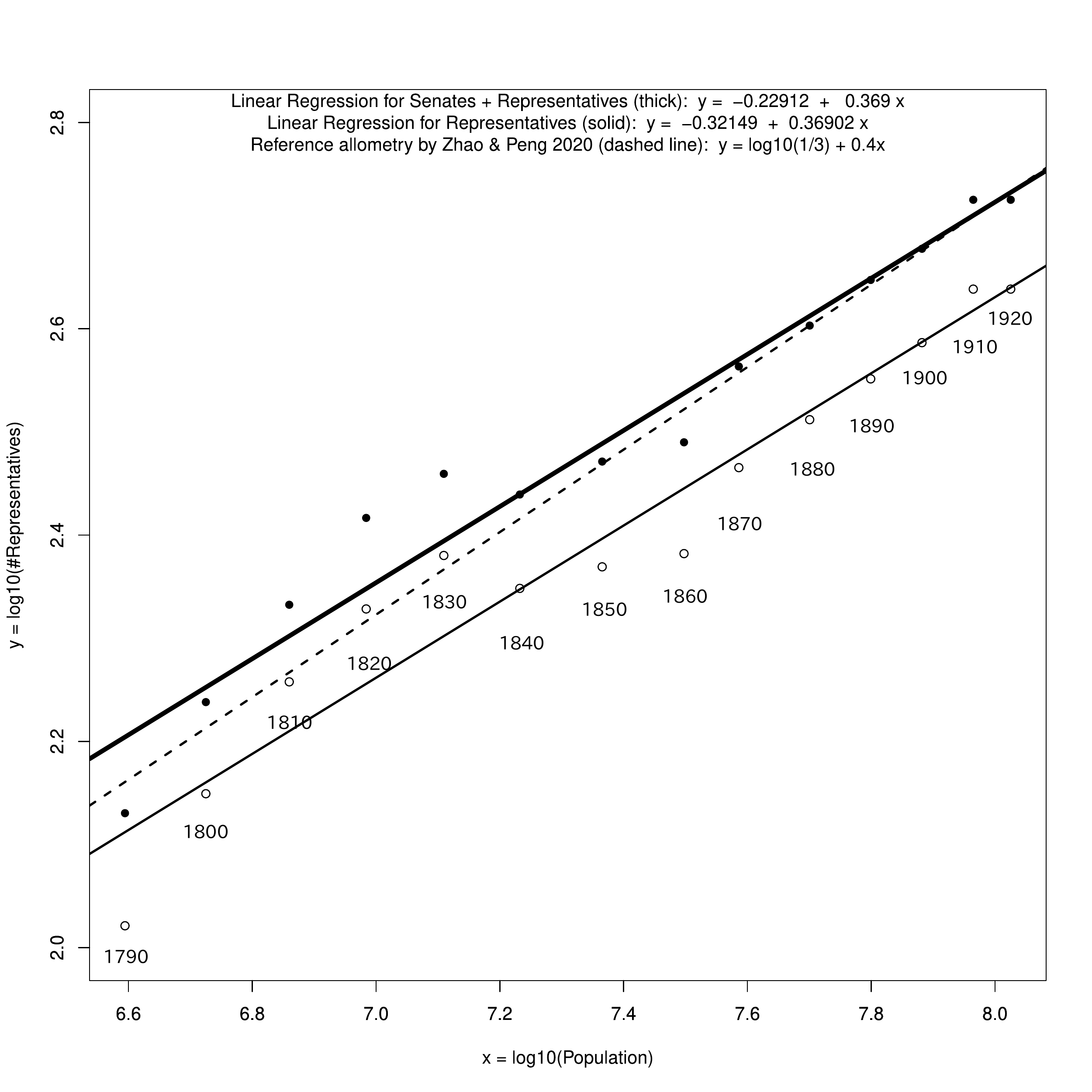}
		\caption{Population of the U.S. (x-axis) and seats of the U.S. Congress (y-axis)
		between 1790 and 1920 in log-scale.
		Black dots denote the size of the Congress (Senate + House),
		whereas white dots denote the size of the House.
		The regression result (thick line) is a p-value of $1.8e-09$ and an adjusted $R^2=0.95$.
		For reference, a standard formula by \cite{zhao_2020} is also plotted (dashed line).
		Data source: \cite{usa-population} for population and \cite{usa-house} for seats.}
		\label{fig:us}
	\end{figure}

	%
	The real-world data suggests $0.4$ as a good choice of $\gamma$ for today's legislatures
	(\cite{auriol_2012,stigler_1976,tanimoto_2022,zhao_2020}; Figure~\ref{fig:us} of this article).
	This phenomenon is surprising: it suggests the existence of an optimal
	number of representatives and that number depends largely on the population,
	little on other factors such as location, race, culture, religion, economics, political system,
	and administrative division (\cite{stigler_1976,taagepera_1972,zhao_2020}).
	We note that voting game and mechanism design models
	 (\cite{auriol_2012,godefroy_2018,ismail_2018,penrose_1946,revel_2022})
	are limited in explaining this phenomenon,
	as they have different values of $\gamma$ and lack connections to {\em representations}.

	On the other hand, social network models consider representation by network connections
	(\cite{taagepera_1972,margaritondo_2021,gamberi_2021,zhao_2020}).
	In particular, the formula $f(p) = \frac{1}{3} \times p^{0.4}$ obtained by
	\cite{zhao_2020} is the first model derived from a theoretical analysis that matches
	the value of $\gamma$ observed in the real-world.
	Surprisingly enough, their formula matches the size of the U.S. Congress
	(Senate + House of Representatives) between 1790 and 1920 with very good fitness (Figure~\ref{fig:us}).
	The existence of social network analysis derived formula supports that the (optimal)
	number of representatives may be largely determined by social connections.
			
	\begin{figure}[htbp]
	\centering
	\includegraphics[width=\textwidth]{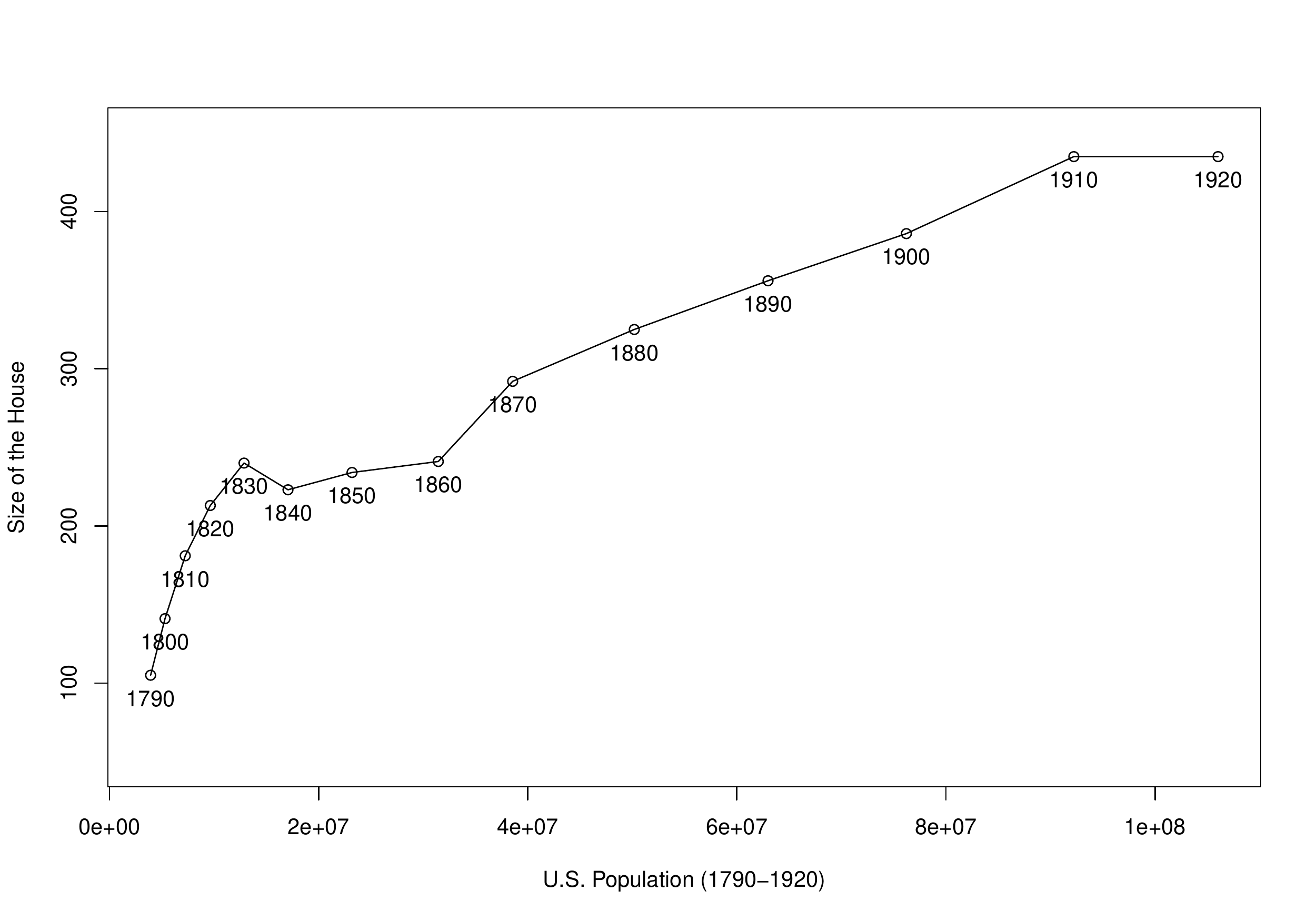}
		\caption{Population of the U.S. and the size of the House between 1790 and 1920 (after 1929, the size
		is fixed to 435). Data sources: \cite{usa-population} and \cite{usa-house}.}
	\label{fig:us_house}
	\end{figure}

	Formula \ref{eqn:general} with $\gamma < 1$ is called {\em subproportional}.
	Subproportionality implies that human society is very efficient in aggregating public voices
	in a {\em superlinear} manner.
	James Madison (1751--1836) was the first to note a subproportionality
	between the number of representatives and the population.
	He proposed an amendment to the U.S. Constitution (\cite{madison_1789})
	that can be considered as a scheme with $\gamma=\frac{1}{2}$ (\cite{blonder_2021}).
	The most interesting idea of Madison's amendment proposal is that it is
	{\em subproportional} in the {\em long-term} but {\em proportional piecewise}
	(\cite{blonder_2021,zhao_2022}).
	This type of subproportionality is considered important, as piecewise ``proportionality''
	is easier to be accepted by the public and implemented than a straightforward nonlinear
	formula shown in (\ref{eqn:general}).
	
	Though Madison's amendment proposal was not adopted, his idea actually has affected
	the size of the House.
	The U.S. Constitution states that the seats in the House must be reapportioned
	to the states according to their populations every ten years (Article I, Section 2, Clause 3).
	It seems that this clause had an implicit understanding
	that the size of the House should increase {\em proportionally} to the total population
	(\cite{madison_1789,kromkowski_1991}),
	which can be approximately observed between 1790 and 1830 (Figure~\ref{fig:us_house}).
	However, the size of the House between 1840 and 1920 increased by substantially
	{\em less} than the size of the population (see Figure~\ref{fig:us_house}).
	Therefore, the size of the House actually followed Madison's piecewise subproportionality proposal.
		
	This piecewise subproportionality scheme went to an extreme in the 1920s,
	when the size of the House was fixed to 435.
	The reason behind this decision was that Congress failed to reapportion seats, because
	they did not have an optimal method for {\em rounding fractional seats to whole numbers}
	(\cite{chafee_1929}).
	%
	%
	Finally, after several failed attempts,
	due to the pressure of the approaching 1930 census,
	Congress ``hastily
	passed reapportionment legislation'' (\cite{kromkowski_1991}, 134) and
	permanently fixed the size of the House, for {\em ease of operation}.

	The discord between the total number of representatives and apportionments
	has led to several issues as discussed by \cite{kromkowski_1991,kromkowski_1992}.
	We note that this discord introduces an inconsistency in the philosophy of the Constitution, namely, that fixing the size of the House suggests that the number
	of representatives is {\em independent} of the population,
	whereas the Constitution states that the number of representatives
	should be proportional to the population.
	Many people have suggested increasing the size of the House
	(\cite{bowen_2021,frederick_2008,frederick_2009,kromkowski_1991,kromkowski_1992,leib_1998,lijphart_1998}).
	%

\subsection{Seat apportionment schemes}\label{subsec:apportion}

	There are three schemes for apportioning a (fixed) number of seats to different subgroups:
	\begin{enumerate}
		\item {\em Fixed Apportionment} (FA): This scheme assigns a fixed number of seats to
			each subgroup.
			An FA scheme is adopted by the U.S. Senate (two seats per state).
			Single and fixed-member district electoral systems are also examples of this type of scheme.
		\item {\em Proportional Apportionment} (PA): This scheme assigns seats proportionally according to
			the populations of subgroups.
			The House and many legislatures worldwide use this type of scheme.
		\item {\em Degressive Proportionality} (DP):
			This is a new type of apportionment scheme that has been adopted by the
			European Parliament (\cite{european_2018}).
			Let us discuss it in the following.
	\end{enumerate}
	As one of the most important steps toward individual equality, PA and PR schemes
	began to dominate
	the literature in the late eighteenth century (\cite{pr}).
	Nevertheless, they have been challenged by DP schemes recently.
	According to the official definition in \cite{european_2018}
	(see also \cite{cegielka_2019,grimmett_2017}), DP is defined as follows.
	For any two subgroups A and B with populations $p_A$ and $p_B$, respectively,
	the numbers of seats $s_A>0$ and $s_B>0$ before rounding to whole
	numbers should satisfy the following constraints:
	\begin{eqnarray}
		\left(\frac{p_A}{s_A} - \frac{p_B}{s_B}\right) \left(p_A - p_B\right) > 0 & \mbox{if $p_A \neq p_B$,}\label{eqn:dp1}\\
		(s_A - s_B)(p_A - p_B) > 0 &  \mbox{if $p_A \neq p_B$,}\label{eqn:dp2}\\
		s_A = s_B & \mbox{otherwise ($p_A = p_B$).}\label{eqn:dp3}
	\end{eqnarray}
	For example, Formula~\ref{eqn:general} is a type of DP when $0<\gamma<1$.
	Note that proportionality requires that $\frac{p_A}{s_A} = \frac{p_B}{s_B}$,
thereby satisfying (\ref{eqn:dp2}) and (\ref{eqn:dp3}) but never satisfying (\ref{eqn:dp1}).
	Hence, despite its name, DP is {\em not} a type of proportionality approach.
	Instead, we propose to use ``subproportionality'' to replace DP.

	The nonproportionality of DP schemes has received criticism for ``unequal'' representations of individuals.
	The European Parliament has explained (see \cite{grimmett_2017}) that DP {\em compromises} between
	{\em individual} equality (per capita) and the equality of {\em state} (per state).
	This ``compromise'' can also be observed in Canada, Germany, and the EU Council (\cite{allen_2017}).
	In fact, the U.S. also utilizes this compromise with its {\em two} chambers:
	The Senate follows a per-state principle, whereas the House follows a per-capita principle.
	By adopting these two principles simultaneously, the U.S. Congress implements a type of
	DP scheme.
	This is why we used the total number in our regression study (Figure~\ref{fig:us}).
	Nevertheless, later we will show that there is actually {\em no compromise}, and the DP method is better than PA/PR approaches in terms of {\em individual} equality.
		
	Finally, we remark on the extensively studied issue
	of an ``optimal'' method for {\em rounding} fractional numbers of seats
	to whole numbers.
	The only consensus is that there is no method that is optimal in terms of all aspects.
	We refer readers to
	\cite{balinski_2001,chafee_1929,huntington_1942,kromkowski_1992,squire_2005} for discussions, and
	\cite{benoit_2000,kalandrakis_2021,puyenbroeck_2008,samuels_2001,taagepera_2003}
	for measuring disproportionality.

\section{Standardizing representation with unbiased inequality}
\label{sec:theory}

	Thus far, we have described how the PSP and PR scheme are population-biased and limited.
	To apportion seats to subgroups equally, we first need to know a standard
	number of representatives for a subgroup (to define the equality).
	In this section, given such a standard function,
	we propose an unbiased indicator to quantify the weight of an individual
	in a subgroup.
	Then, we use this indicator to derive an apportionment scheme with unbiased individual equality.
		
	Suppose that a function $f = f(p)$ for the standard
	number of representatives (seats) for population $p$ is available.
	Such a standard function can be determined according to the average number
	obtained by an empirical study, or the optimal number obtained by a theoretical study,
	or a model adopted by policy makers (e.g., the formula adopted by the European Parliament).
	See Subsection~\ref{subsec:number} for examples.

	We first assume that $f$ is invertible.
	This assumption is true for all existing models (Table~\ref{tab:number})
	except for the $\gamma=0$ case.
	Let $f^{-1}$ denote the inverse function of $f$.
	Suppose that $s$ seats are assigned to population $p$.
	For estimating the contribution (weight) of an individual,
	we propose to use the ratio $\frac{p^*}{p}$ of the
	{\em effective} population $p^*$ to the {\em real} population $p$,
	where the effective population $p^*$ is the standard population that deserves $s$ seats,
	i.e., $p^* = f^{-1}(s)$.
	Therefore, the proposed {\em population seat index} (PSI) is as follows.
	\begin{equation}\label{eqn:main}
		w(s, p) \ = \ \frac{f^{-1}(s)}{p}.
	\end{equation}
	Note that (\ref{eqn:main}) calculates a {\em scalar}.
	This value is equal to $1$ if $s=f(p)$ is the standard number of seats for population $p$.
	Thus, we can estimate the {\em absolute inequality} of an assignment
	with respect to the standard value.
	If the value calculated by Formula \ref{eqn:main} is greater than 1,
	the number of seats $s$ is greater than the standard number, i.e.,
	overrepresentation; however, if the value is less than 1, the number
	of seats $s$ is less than the standard number, i.e., underrepresentation.
	This fact is independent of the population $p$.
	Therefore, the proposed PSI indicator has no population-dependent bias that the PSP has.

	We illustrate this concept with the example in Section~\ref{sec:introduction},
	where $f(p)= \sqrt{p}$ (see Figure~\ref{fig:illustration}).
	Formula \ref{eqn:main} shows $w(s,p)=\frac{s^2}{p}$.
	Suppose that there are 5 seats and we assign 2 to Group A (9 people) and 3 to Group B (16 people),
	respectively.
	The (unbiased) weight of an individual
	in Group A is thus $w(2, 9) = \frac{2^2}{9} = \frac{4}{9}$,
	less than the weight $w(3, 16)=\frac{9}{16}$ of Group B.
	Therefore, people in Group A are {\em less} represented than people in Group B,
	in contrast to the (biased) analysis according to the PSP which calculates
	$\frac{2}{9} > \frac{3}{16}$.
	In fact, we can determine an optimal (fractional)
	apportionment with {\em relative} equality by solving an equation $w(s_A, p_A) = w(s_B, p_B)$,
	where $s_A + s_B = 5$, $p_A=9$, and $p_B=15$.
	A simple calculation shows that $s_A = 15/7$ and $s_B = 20/7$.
	Rounding these values to the nearest whole numbers, we obtain $2$ and $3$ seats, respectively.
	This apportionment gives less representation to Group A (as $15/7 > 2$)
	and more representation to Group B (as $20/7<3$), matching the above analysis.
	Note that the absolute equality, i.e., $w(s, p) = 1$ for all groups, occurs
	if and only if the total number of seats is $3+4=7$.
		
	In general, we can use the PSI to determine apportionments with absolute or relative individual equality.
	Assume that there are $k$ groups with populations $p_1, p_2, \ldots, p_k$.
	Given a total number $S$ of seats, the fractional apportionment problem with (unbiased)
	{\em relative} equal weight $w^*$ can be formulated as determining the number
	$s_i$ of seats assigned to group $i$, where $i=1, 2, \ldots, k$, as follows:
	\begin{eqnarray}
		w^* = \frac{f^{-1}(s_1)}{p_1} = \frac{f^{-1}(s_2)}{p_2} = \cdots = \frac{f^{-1}(s_k)}{p_k},\label{eqn:equation1}\\
		s_1 + s_2 + \cdots + s_k = S.\label{eqn:equation2}
	\end{eqnarray}
	According to (\ref{eqn:equation1}), we have
	\begin{equation}\label{eqn:solution}
		s_i = f(w^* p_i) \ \mbox{ for $i=1,2,\ldots,k$}.
	\end{equation}
	Then, according to (\ref{eqn:equation2}), the weight $w^*$ can be calculated by solving:
	\begin{equation}\label{eqn:weight}
		\sum_{i=1}^{k}f(w^* p_i) \ = \ S.
	\end{equation}
	Once $w^*$ is found, the fractional apportionment of seats can be determined with
	Formula~\ref{eqn:solution}.
	We note that this scheme includes the traditional PR scheme as a special case
	when the standard function $f(p) \propto p$.
	%
	%
	An integer apportionment can then be obtained by some rounding method.

	The solution of Formula~\ref{eqn:weight} depends on $f$.
	We consider $f(p) = a + bp^\gamma$ for some constants $a, b, \gamma$
	with $\gamma \neq 0$ (in practice, $a$ is used to fit a minimal number of seats).
	The proposed indicator PSI is then
	\begin{equation}\label{eqn:simple}
		w(s, p) \ = \ \frac{((s-a)/b)^{1/\gamma}}{p}.
	\end{equation}
	The constant $(1/b)^{1/\gamma}$ can be removed if we are interested in only
	{\em relative} equality.
	%
	%
	Then we can see that the $a=0, \gamma=1$ case degenerates to the
	PSP $\frac{s}{p}$.
	For the general case, we can derive an apportionment scheme
	with individual equality as follows.
	A simple calculation with (\ref{eqn:weight}) shows that the weight is
	\begin{equation}\label{eqn:weight2}
		w^*  \ = \ \left(\frac{S-ka}{b \sum_{i=1}^{k} p_i^\gamma}\right)^{1/\gamma}.
	\end{equation}
	Therefore, according to (\ref{eqn:solution}), the number of seats (with relative equality)
	can be calculated as
	\begin{equation}\label{eqn:solution2}
		s_i = a + \frac{p_i^\gamma}{\sum_{j=1}^{k} p_j^\gamma} \times (S-ka) \ \mbox{ for $i=1,2,\ldots,k$}.
	\end{equation}
	When $a=0$, this is to distribute seats proportionally to the
	$\gamma$-th powers of the populations.
	We remark that, for absolute equality, i.e., $w^*=1$,
	the total number of seats must be $S = ka + b\sum_{j=1}^{k} p_j^\gamma$.

	Finally, let us consider a function $f$ with no inverse function.
	We only consider $f = a$ for a constant $a>0$, which is adopted by the U.S. Senate.
	Let $f_\epsilon(p) = a + \epsilon p$,
	where $\epsilon > 0$ is a small number.
	We define the weight $w(s, p)$ for $f$ according to the limit of $w_\epsilon(s, p)$
	when $\epsilon \to 0$:
	\begin{eqnarray}
		w_\epsilon(s, p) = \frac{s-a}{\epsilon p} \to \left\{\begin{array}{lc}
			0, & s=a,\ \epsilon \to 0,\\
			+\infty, & s>a,\ \epsilon \to 0,\\
			-\infty, & s<a,\ \epsilon \to 0.
		\end{array}\right.
	\end{eqnarray}
	The weight is $w(s, p)=0$ if $s=a$.
	This result matches our intuition, as no one contributes to the number of seats.
	Otherwise, if $s>a$ (respectively, $s<a$), the weight is $+\infty$ ($-\infty$),
	which we consider is reasonable.
	In either case, the result is independent of the population.
	The only equal apportionment is $s_i = a$ for all $i$ (thus, $S = ka$).
	Therefore, we conclude that the apportionment of the U.S. Senate is consistent with respect to
	individual equality, where each individual has a constant contribution (weight) of zero.

\section{Implications}
\label{sec:discussion}

\begin{figure}[htbp]
	\centering
		 \includegraphics[width=0.9\textwidth]{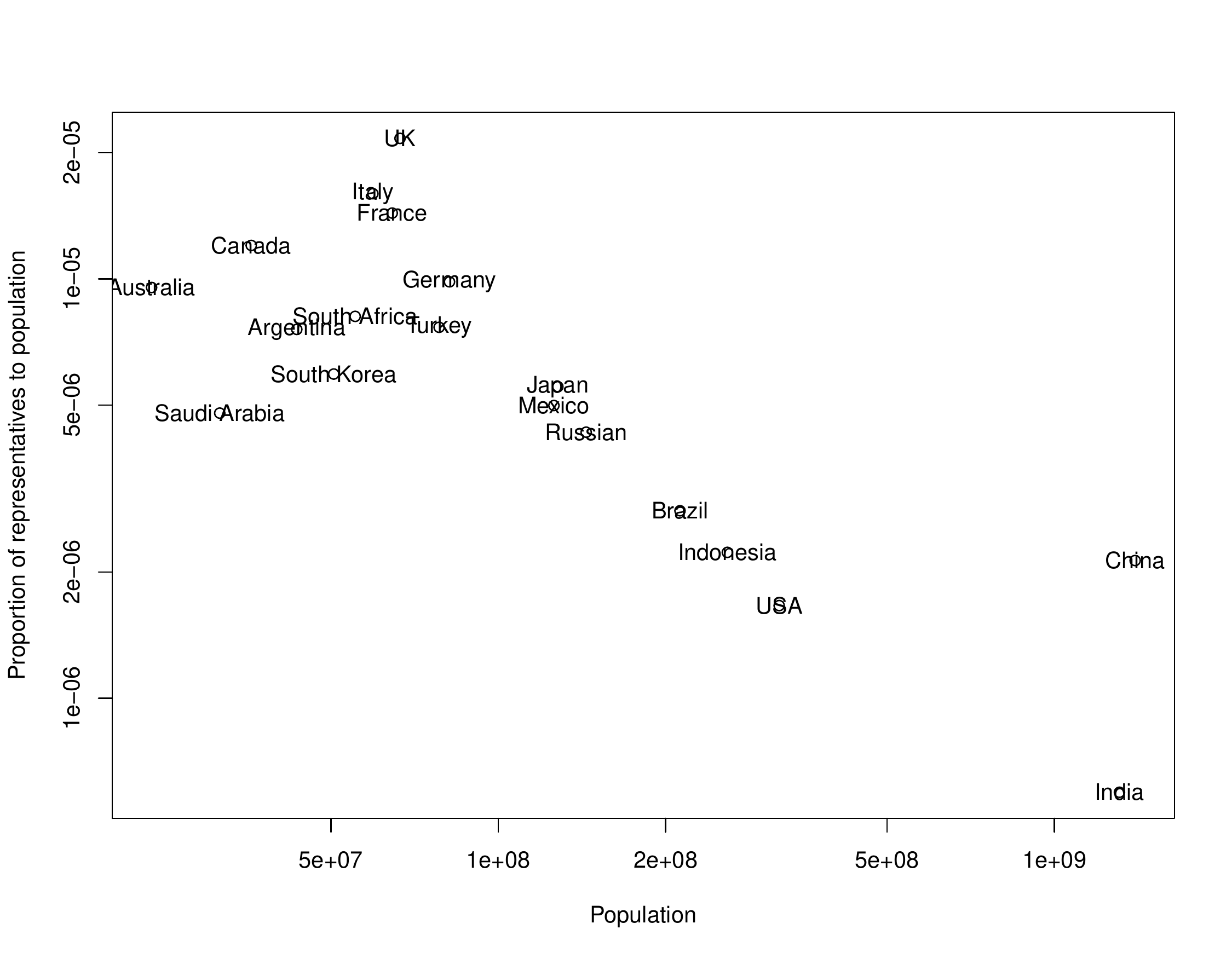}
		 \includegraphics[width=0.9\textwidth]{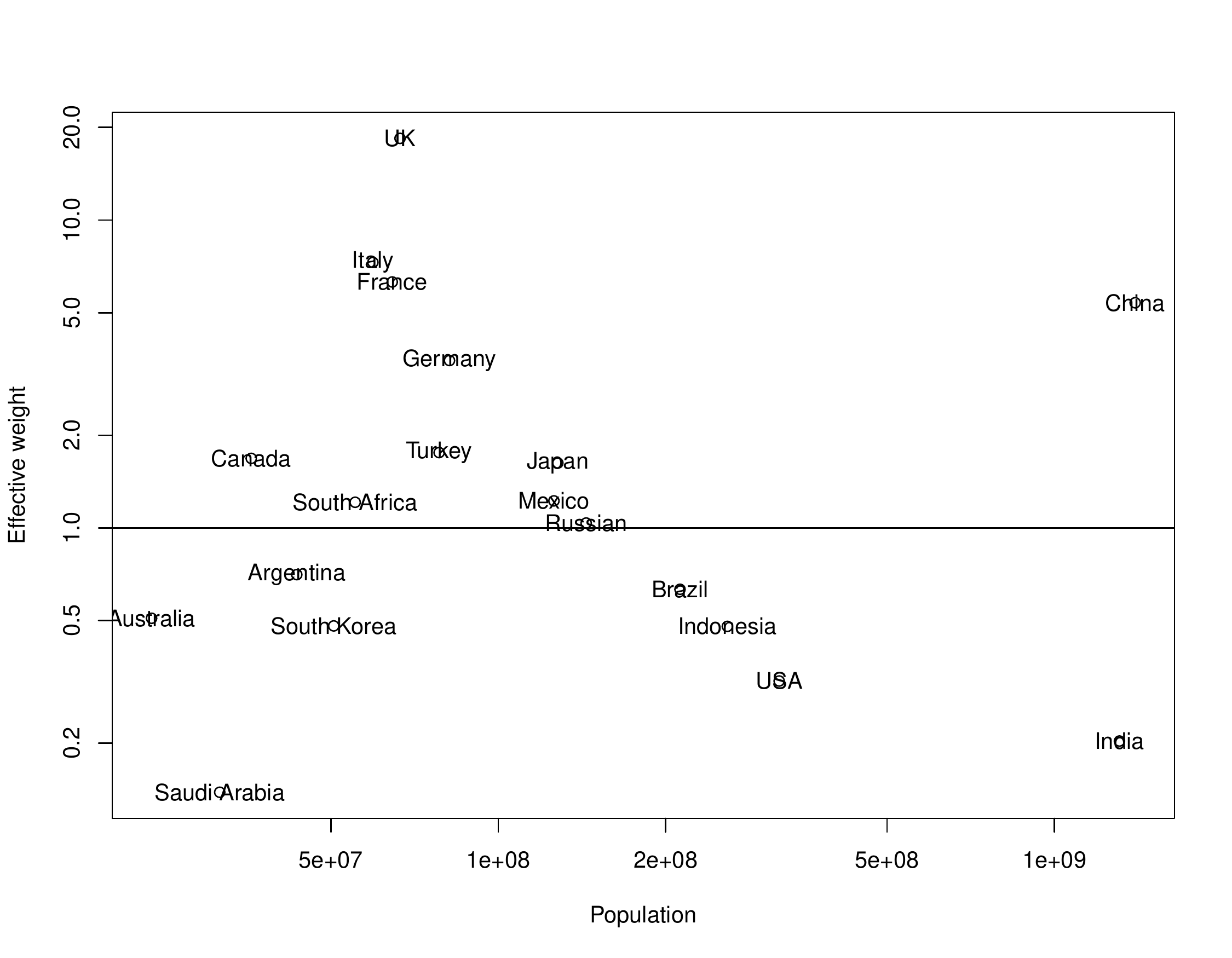}
		\caption{Comparing G20 countries (except for the EU) according to the indicators
		PSP (upper) and PSI (lower), where the total size of the congress is used for a bicameral country.
		For the PSI, $f(p) = \frac{1}{3}p^{0.4}$ (\cite{zhao_2020}) is used as
		the standard function for the global average.
		Note that the PSP provides no measure of the appropriateness,
		while the proposed PSI does: ${\rm PSI}=1$ indicates absolute compatibility
		with the global average; ${\rm PSI}>1$ (respectively, ${\rm PSI}<1$)
		indicates that the size of the congress is more (less) than the global average.
		Differences in the order due to different indicators can be confirmed,
		e.g., Canada and Germany, Saudi Arabia and the USA, etc.
		Additionally, Australia and Germany have almost the same PSP value but considerably different PSI values.
		Data source: \cite{ipu_2021}.
		}
	\label{fig:unbiased_g20}
	\end{figure}

	We discuss the implications of the proposed theory for existing studies.
	First, we empirically compare the PSI with the PSP using data from G20 countries (except for the EU).
	For the standard function, we used $f(p) = \frac{1}{3}p^{0.4}$ (\cite{zhao_2020}), as
	this function shows the average size of the congress in the world through regression.
	In other words, our study compares G20 countries with the average of the world.
	Figure~\ref{fig:unbiased_g20} shows the results,
	where ``Effective weight'' denotes the PSI calculated by Formula~\ref{eqn:main}.
	We noted several interesting findings in the figure.

	Next, we consider the theoretical implications for existing apportioning schemes.
	Given a standard function, Section~\ref{sec:theory} has given the scheme to apportion seats
	with equality.
	However, what would happen if a different scheme is used?
	To study this question, let us consider a standard function
	$f_1(p)=a_1 + b_1 p^{\gamma_1}$
	and an apportioning function $f_2(p)=a_2 + b_2 p^{\gamma_2}$,
	where $a_i,b_i,\gamma_i$ are nonnegative constants and $\gamma_i \le 1$, $i=1,2$.
	Besides, we use $p_1$ and $p_2$ to denote the populations
	of two groups to study relative equality.
	
	First of all, notice that, for a fixed-member district electoral system (i.e., the total
	number of seats is a constant to all electoral districts),
	regardless of how the functions $f_i(p)$ are chosen,
	relative equality is always satisfied if $p_1=p_2$.
	Thus, our theory is compatible with equal redistricting (see \cite{auerbach_1964}).
	For the general case, we summarize the result in Table~\ref{tab:implication}.

%

	\begin{table}[htb]
		\caption{Study on the equality if an apportioning scheme $f_2(p)=a_2+b_2p^{\gamma_2}$ is used.}\label{tab:implication}
	\centering
	\begin{tabular}{c|c|c|c}
		{Standard function $f(p)$} & PSI & {For absolute equality}  & {For relative equality} \\\hline
		$a_1+b_1 p^{\gamma_1}$ & $\frac{(a_2+b_2p^{\gamma_2}-a_1)^{1/\gamma_1}}{b_1^{1/\gamma_1}p}$ & $\frac{a_2+b_2p^{\gamma_2}-a_1}{b_1p^{\gamma_1}}=1$ & $\frac{(a_2+b_2p_1^{\gamma_2}-a_1)p_2^{\gamma_1}}{(a_2+b_2p_2^{\gamma_2}-a_1)p_1^{\gamma_1}}=1$
	\end{tabular}
	\end{table}

	From the above result, it can be observed that when $f_2(p)=f_1(p)$, absolute (hence relative)
	equality can always be guaranteed.
	Otherwise, only in special cases the equality can be obtained.
	In particular, since $f_1(p) \propto p^{\gamma_1}$ for some $0<\gamma_1<1$ in the real world,
	the PA (i.e., the PR, where $\gamma_2 = 1 > \gamma_1$) scheme ensures (relative) equality
	only when $p_1=p_2$, i.e., equal redistricting.
		
	We remark that the PA scheme adopted by the House and various
	legislatures worldwide is not truly proportional.
	A general idea of such an approach is to first determine a population size $d$,
	which should ideally be $d = \frac{\sum_j p_j}{S}$ for a total of $S$ seats,
	then, assign $\frac{p_i}{d}$ seats (before rounding to whole numbers) to group $i$
	with population $p_i$ for all $i$.
	However, in general, $d$ is not a constant; thus,
	$\frac{p_i}{d} = \frac{S p_i}{\sum_j p_j}$ is not truly proportional to $p_i$.
	As previously discussed, this approach has a population-dependent bias and thus
	cannot ensure representation equality among individuals.
	To achieve true equality, we should either resize the total number of
	seats in proportional with the total population, as the House did in 1790--1830,
	or adopt a subproportional apportionment scheme.

\section{Conclusion and Discussion}
\label{sec:conclusion}

	In the previous century, the literature has noted the existence of a standard
	(average or optimal) number of representatives (seats) for a population,
	and that follows some subproportionality scheme with respect to the size of the population.
	Based on these previous works, this article pointed out a bias inherent in the PSP metric
	and thus in the PR scheme in estimating the contribution (weight) of an individual,
	which implies that the PR is insufficient to ensuring representation equality of individuals.
	To address this issue, we introduced a generalized apportionment problem with a
	standard function $f(p)$ for the number of seats
	for a population $p$ and a nonproportional indicator PSI.
		
	It is shown that the proposed indicator does not have the bias inherent to the PSP.
	By using it as the indicator to develop an apportionment scheme,
	a standardized representation theory is proposed with absolute or relative individual equality.
	In particular, if $f(p) \propto p^\gamma$ for some constant $\gamma$,
	the proposed scheme distributes seats proportionally to the
	$\gamma$-th power of the populations.
	Because $0 \le \gamma < 1$ in the real world,
	it is concluded that the PR scheme represents people in smaller groups {\em less}
	than people in larger groups, whereas the proposed subproportionality scheme,
	which is a type of degressive proportionality approach, guarantees equality.
		
	This article is limited in three aspects.
	Firstly, it only considered an ideal fractional solution in theory.
	In the future, rounding methods must be provided in order to apply the theory in practice.
	We note that existing rounding methods need to be revised since they
	were designed for PR schemes only.
	Secondly, to support a subproportional apportioning scheme, more studies are required to
	investigate a standard size of {\em subnational} legislatures.
	Lastly, the most importantly, it lacks discussion on how to decide an appropriate standard
	function.
	Future works may study this issue with the EU Parliament.

\section*{Funding}
This work was supported by the Japan Society for the Promotion of Science (JSPS),
Grant-in-Aid for Scientific Research (C) [JP8K11182].

\section*{Acknowledgments}
We thank Prof. Takashi Sekiyama for the valuable discussions.


\printbibliography

\end{document}